\documentclass[12pt,letter]{article}

\usepackage[margin=.8in]{geometry}

\usepackage{amssymb}
\usepackage{filecontents,lipsum}
\usepackage[noadjust]{cite}
\usepackage[percent]{overpic}
\usepackage{stfloats}
\usepackage{ctable}
\usepackage[mathlines]{lineno}

\usepackage{cite}
\usepackage{graphicx}
\usepackage{caption}
\usepackage{subcaption}
\usepackage{authblk}
\usepackage{xcolor}
\usepackage{amsmath}
\usepackage{array}
\usepackage[normalem]{ulem}

\begin{document}
\date{}
\title{Saving Moore's Law Down To 1nm Channels With Anisotropic Effective Mass} 

\author[1*]{{Hesameddin Ilatikhameneh}}
\author[1*]{{Tarek Ameen}}
\author[1]{{Bozidar Novakovic}}
\author[1]{{Yaohua Tan}}
\author[1]{Gerhard Klimeck}
\author[1]{Rajib Rahman}

\affil[1]{Department of Electrical and Computer Engineering, Purdue University, USA}
\affil[*]{\normalsize{These authors contributed equally to this work.}}

\maketitle
\setlength{\textfloatsep}{7pt plus 1.0pt minus 2.0pt}
\providecommand{\keywords}[1]{\textbf{\textit{Keywords---}} #1}

\begin{abstract}
Scaling transistors' dimensions has been the thrust for the semiconductor industry in the last 4 decades. However, scaling channel lengths beyond 10 nm has become exceptionally challenging due to the direct tunneling between source and drain which degrades gate control, switching functionality, and worsens power dissipation. Fortunately, the emergence of novel classes of materials with exotic properties in recent times has opened up new avenues in device design. Here, we show that by using channel materials with an anisotropic effective mass, the channel can be scaled down to 1nm and still provide an excellent switching performance in both MOSFETs and TFETs. In the case of TFETs, a novel design has been proposed to take advantage of anisotropic mass in both ON- and OFF-state of the TFETs. Full-band atomistic quantum transport simulations of phosphorene nanoribbon MOSFETs and TFETs based on the new design have been performed as a proof.

\end{abstract}

Shrinking the size of metal oxide semiconductor field effect transistors (MOSFETs) has improved the functionality, speed, and cost of microprocessors over the last four decades. However, the advantages of scaling are quickly fading away \cite{Ionescu}. For example, the operational frequency of CPUs has stopped improving since 2003 due to power consumption of CPUs reaching their cooling limit ($\approx$100W/cm$^2$) \cite{BeyondCMOS}. Moreover, scaling down $L_{ch}$ towards the few nanometer regime is becoming more challenging due to source-to-drain (SD) leakage current \cite{Sub12, Lundstrom_SD}; the gate controlled potential barrier becomes more transparent as channel becomes shorter and direct SD tunneling increases. Another challenge in miniaturizing MOSFETs is scaling down the supply voltage $V_{DD}$ \cite{BeyondCMOS}. A smaller $V_{DD}$ can be achieved in a switch with sharper ON to OFF transition. However, the steepness of conventional MOSFETs have a fundamental limit due to thermionic injection of carriers over the channel barrier (60 $mV/decade$ at room temperature). Accordingly, $V_{DD}$ in MOSFETs does not scale very well. On the other hand, tunnel FETs (TFETs) can, in principle, provide steeper switching \cite{Appenzeller1, Appenzeller2}. Nevertheless, scaling TFETs is even trickier than MOSFETs, since scaling affects both ON- and OFF-states of the TFETs \cite{Sub10, Ian, Seabaugh}. Hence, the tremendous improvement in processing power of transistors every few years linked to the dimension scaling and empirically described by Moore's law has reached a dead end. Fortunately, it is shown here that 2D materials with anisotropic effective mass ($m^*$) can be used to solve these problems and save Moore's law.

\begin{figure}[!t]
        \centering    
        \begin{subfigure}[b]{0.6\textwidth}
               \includegraphics[width=\textwidth]{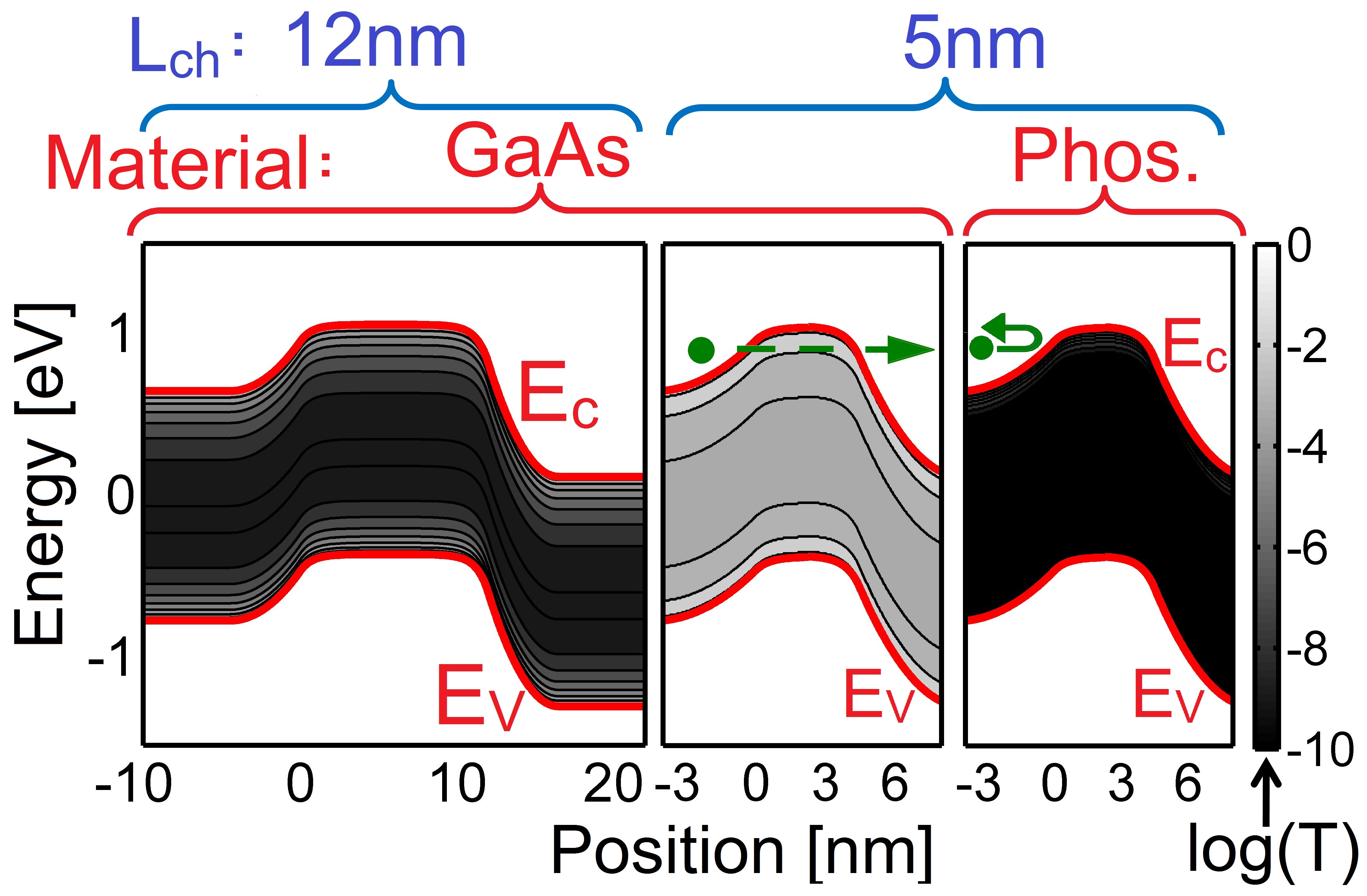}
               \vspace{-1.3\baselineskip}
                \label{fig:Same_EoT}
        \end{subfigure}%
        \vspace{-0.3\baselineskip}
        \caption{The band diagram of a) 12nm long GaAs, b) 5nm long GaAs, and c) 5nm long phosphorene MOSFETs. The colormap shows the transparency of the channel. The potential barrier in the 5nm long GaAs MOSFET is transparent and hence, the gate efficiency is low. This problem can be solved by using phosphorene with high $m^*$.}\label{fig:Fig1}
\end{figure}


First, we discuss the source-to-drain tunneling challenge of the ultra scaled MOSFETs. Reducing the channel size makes the potential barrier more transparent. To visualize this, the transmission is shown in colormap on a logarithm scale and with an overlayed band diagram of MOSFETs in Fig. \ref{fig:Fig1}. The band diagram and transmission profile of a 12nm and 5nm long channel GaAs MOSFET are compared respectively in Fig. \ref{fig:Fig1}a and \ref{fig:Fig1}b. 5nm long channel GaAs MOSFET suffers significantly from SD leakage which reduces the gate control. Equation (\ref{eq:tun1}) shows the dependence of tunneling current through barrier on $m^*$ of the channel material. According to Equ. (\ref{eq:tun1}), an apparent solution to the high transparency of channel barriers in short channel regime is a channel material with higher effective mass.
\begin{equation}
\label{eq:tun1}
log(I_{OFF}) \propto L_{ch} \sqrt{m^*}
\end{equation}
Although high $m^*$ channel materials block SD tunneling effectively, they have a set of drawbacks too. Quantum capacitance ($C_Q$) of channel material increases as a result of larger density of states (DOS) and $m^*$. Accordingly, the gate capacitance ($C_G$) which is the net series capacitance of $C_Q$ and oxide capacitance ($C_{ox}$) increases. Hence, a larger $m^*$ translates into a larger switching delay ($\tau = C_G V_{DD} / I_{ON}$). 
\begin{equation}
\label{eq:Cg}
\frac{1}{C_G} = \frac{1}{C_{ox}} + \frac{1}{C_Q}
\end{equation}
Anisotropic effective mass can provide a solution to this problem with reducing $C_Q$ by a factor of $\sqrt{m^*_{l}/m^*_{h}}$. This reduction of $C_Q$ is the result of the decreased density of states ($DOS$) in anisotropic materials:
\begin{equation}
\label{eq:Cq}
C_Q = q^2 DOS = q^2 \frac{\sqrt{m^*_{l} m^*_{h}}}{\pi \hbar^2}
\end{equation}
\noindent where $m^*_{l}$ and $m^*_{h}$ are low and high effective masses of the channel material along its two main axes. If high $m^*$ axis of channel is aligned with transport direction and low $m^*$ axis is aligned with the confinement direction, both low transparency and small switching delay can be achieved. Note that high $m^*$ along the channel increases the carriers decay rate through barrier exponentially, whereas low confinement $m^*$ reduces DOS and $C_Q$. Hence, a 2D material such as phosphorene \cite{B0} with anisotropic $m^*$ \cite{B1} can provide an excellent switching performance in MOSFETs ensuring the continuation of Moore's Law to atomic dimensions. 

\begin{figure}[!t]
        \centering    
        \begin{subfigure}[b]{0.46\textwidth}
               \includegraphics[width=\textwidth]{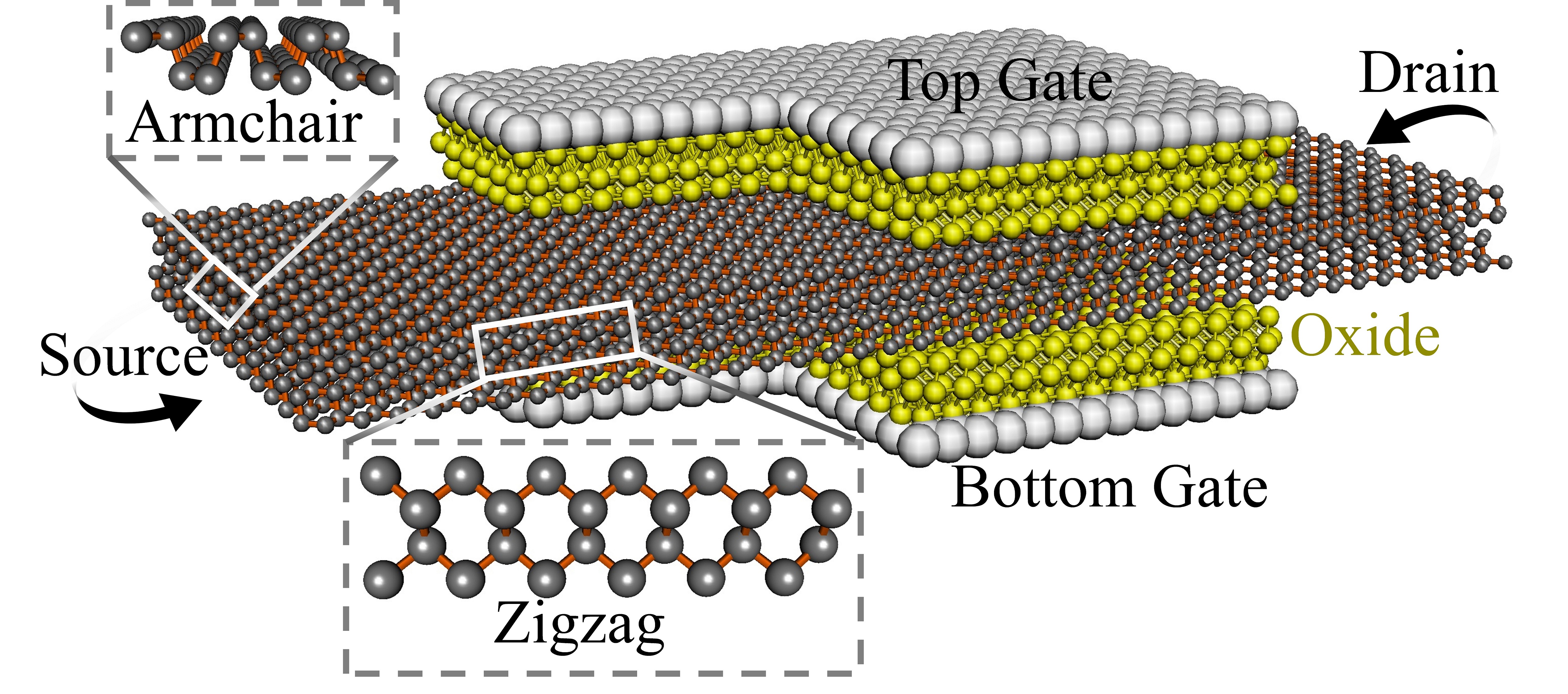}
               \vspace{-1.55\baselineskip}
                \caption{}
                \label{fig:Spacing}
        \end{subfigure}%
        \begin{subfigure}[b]{0.36\textwidth}
               \includegraphics[width=\textwidth]{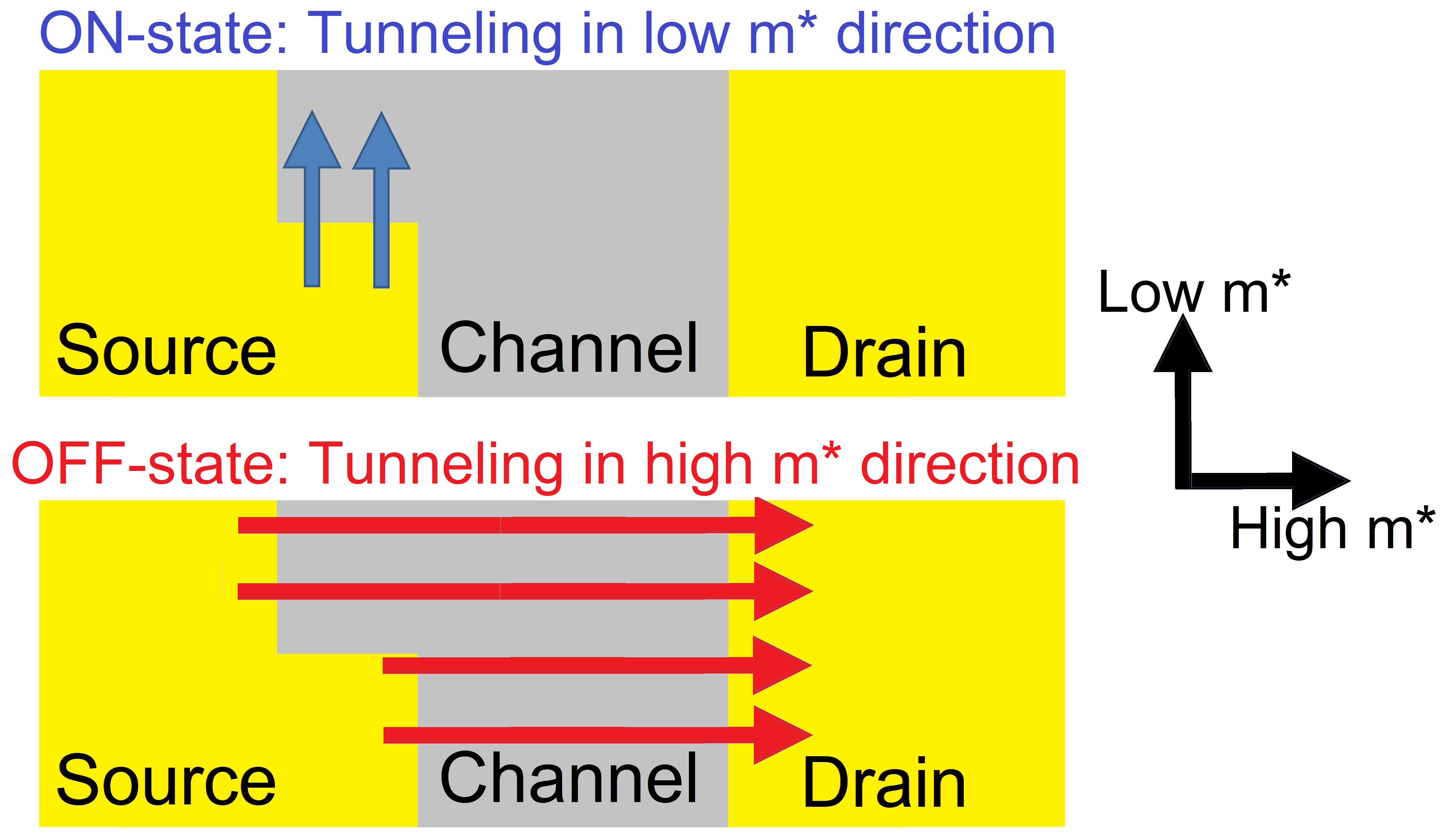} 
               \vspace{-1.45\baselineskip}
                \caption{}
                \label{fig:struct1}
        \end{subfigure}%
        \vspace{-0.5\baselineskip}
        \caption{a) The device structure of the phosphorene TFET with L-shaped gate. b) The main tunneling paths in the ON-state (blue arrows) and OFF-state (red arrows) of the phosphorene TFET.}\label{fig:Fig2}
\end{figure}

Here, we discuss the scaling challenge of TFETs. Although TFETs were intended to reduce the power consumption of transistors \cite{Appenzeller1, Appenzeller2}, scaling TFETs below 10nm is even more challenging than MOSFETs \cite{Ian, Seabaugh, Sub10}. The ON-state and OFF-state tunneling currents (I$_{\rm ON}$ and I$_{\rm OFF}$) depend on the same device parameters \cite{Analytic2}. Thus decreasing I$_{\rm OFF}$ would reduce I$_{\rm ON}$. Roughly, the ON/OFF ratio of TFETs depends on \cite{Sub10, Analytic2, Kane}:
\begin{equation}
\label{eq:tun2}
\frac{log(I_{\rm ON})}{log(I_{\rm OFF})} \propto \frac{L_{ch}}{\Lambda}  \frac{\sqrt{m^*_{r1} E_{g1}}} { \sqrt{m^*_{r2} E_{g2}}}
\end{equation}
\noindent where $\Lambda$ and $L_{ch}$ are the tunneling distances in the ON- and OFF-state respectively. $m^*_{r1}$ and $E_{g1}$ ($m^*_{r2}$ and $E_{g2}$) are the reduced effective mass and the bandgap of the channel material (source-to-channel junction), respectively. 

Shrinking the channel length to few nanometers brings $L_{ch}/\Lambda$ close to 1 and reduces I$_{\rm ON}$/I$_{\rm OFF}$ significantly. One apparent solution can be a heterostructure channel where the term $m^*_{r2} E_{g2}$ is much smaller than $m^*_{r1} E_{g1}$ due to different materials used in the source and channel regions \cite{Wenjun, Het1}. However, heterostructure TFETs suffer from interface states which deteriorate their OFF-state performance \cite{Het2, Takagi, Sapan}. Although homojunction TFETs do not have the interface states, it is challenging to provide high ON/OFF ratio especially below 6nm \cite{Sub10, Tarek1}. Anisotropic effective mass can also provide a solution for this challenge by setting source-channel junction along low $m^*$ axis of channel material and the channel barrier along high $m^*$ axis. 

Although, many novel materials and designs have been proposed to enhance the performance of TFETs such as 2D material TFETs \cite{Hesam1, Fiori, Fan}, Nitride heterostructures \cite{Wenjun}, dielectric engineering \cite{Hesam3}, there are not many proposals for solving the scaling challenge of TFETs \cite{Sub10}. In this work, a new TFET design is proposed to overcome the scaling challenge and enable downsizing to 2nm channel lengths. Fig. \ref{fig:Fig2}a shows a novel TFET device structure to take advantage of anisotropic effective mass. Notice that the gate is L-shaped. Fig. \ref{fig:Fig2}b depicts that the tunneling in the ON-state occurs along the low $m^*$ axis of the channel enhancing the I$_{\rm ON}$. 
However, the tunneling in the OFF-state occurs along the high $m^*$ axis and results in a very low I$_{\rm OFF}$. Hence, this new TFET design can revive Moore's law for sub-10nm TFETs.

In this work, phosphorene nanoribbon has been chosen as the channel material since it has a large effective mass anisotropy in zigzag and armchair directions. Moreover, multi-layer phosphorene provides a range of bandgap ($E_g\approx$1.45 to 0.4eV \cite{B2, B3}) suitable for transistor applications. $E_g$ of monolayer (1L-) and bilayer (2L-) phosphorene is about 1.45eV and 0.8eV, respectively. Since MOSFETs require larger $E_g$ for a smaller source-to-drain leakage, a monolayer phosphorene has been used here. The situation is more tricky in TFETs which need optimized $E_g$. It was shown previously that 2L-phosphorene has an optimum $E_g$, and hence 2L has been chosen for TFETs. HfO$_2$ is used as the gate dielectric with an equivalent oxide thickness (EOT) of 0.5nm in both MOSFETs and TFETs and I$_{\rm OFF}$ is set to $10^{-4} \mu A/\mu m$.

\begin{figure}[!t]
        \centering
        \begin{subfigure}[b]{0.25\textwidth}
               \includegraphics[width=\textwidth]{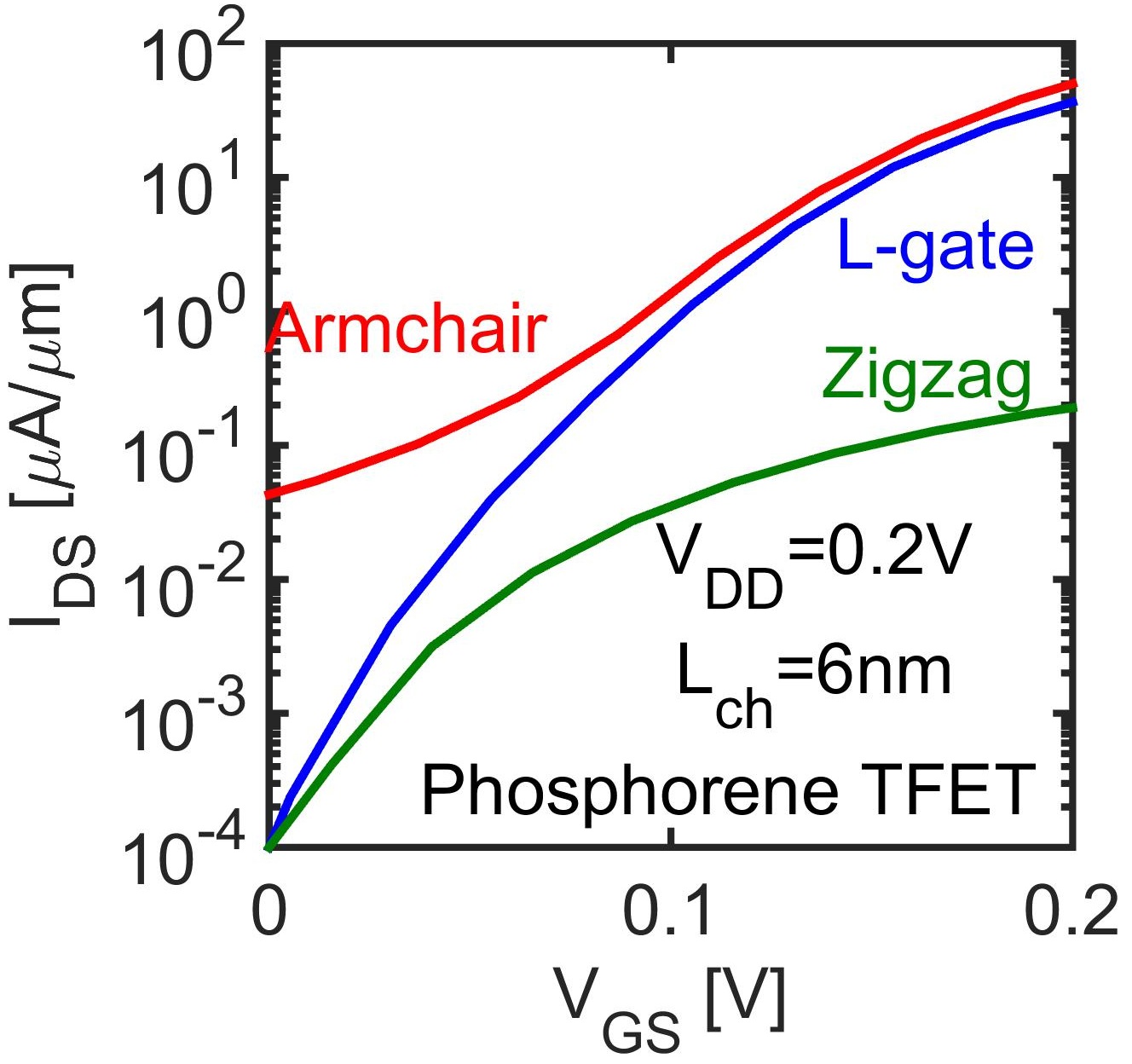} 
               \vspace{-1.55\baselineskip}               
                \caption{}
                \label{fig:laplace}
        \end{subfigure}%
        \begin{subfigure}[b]{0.25\textwidth}
               \includegraphics[width=\textwidth]{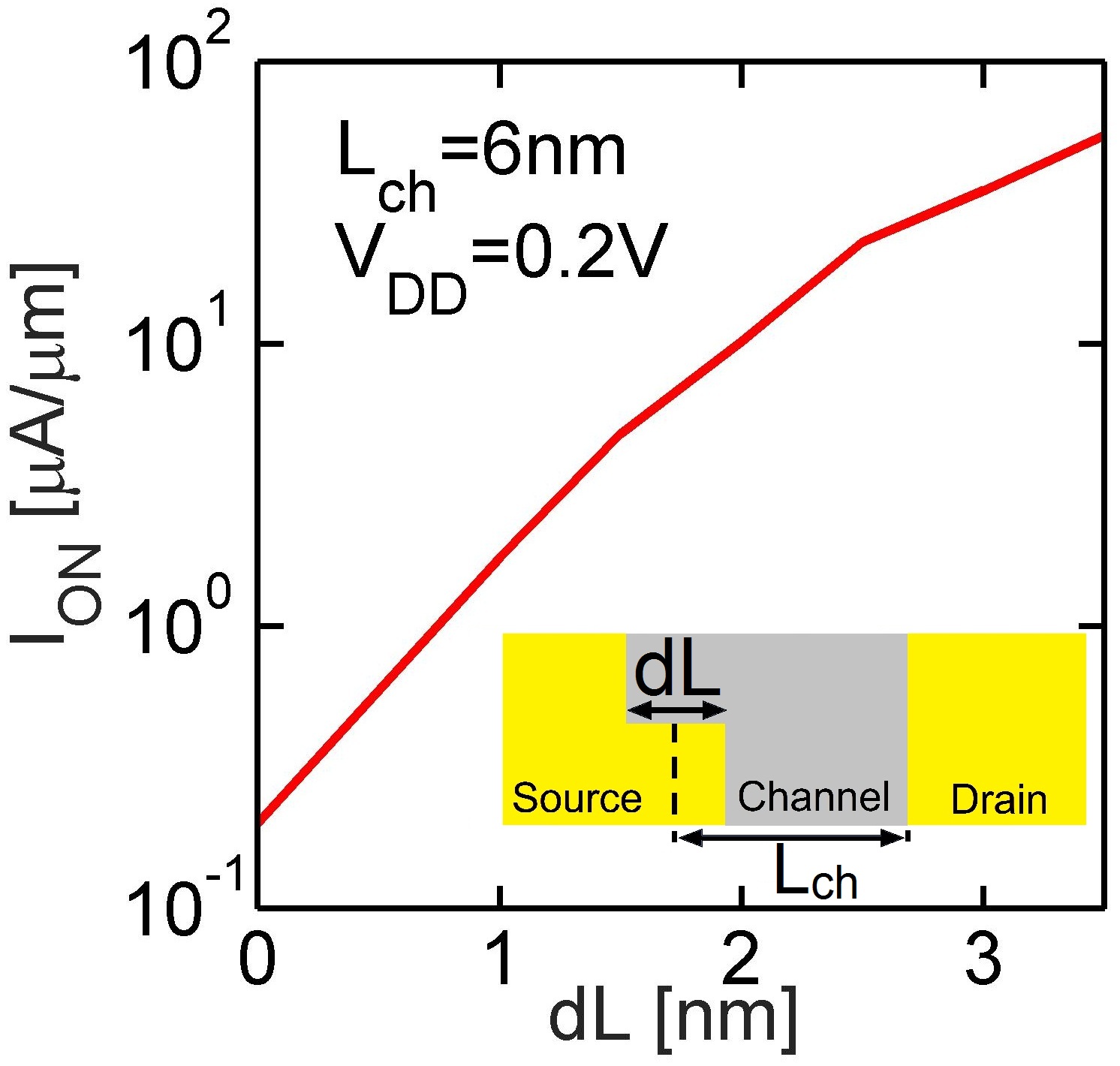}
               \vspace{-1.5\baselineskip}               
                \caption{}
                \label{fig:laplace}
        \end{subfigure}%
                \begin{subfigure}[b]{0.25\textwidth}
               \includegraphics[width=\textwidth]{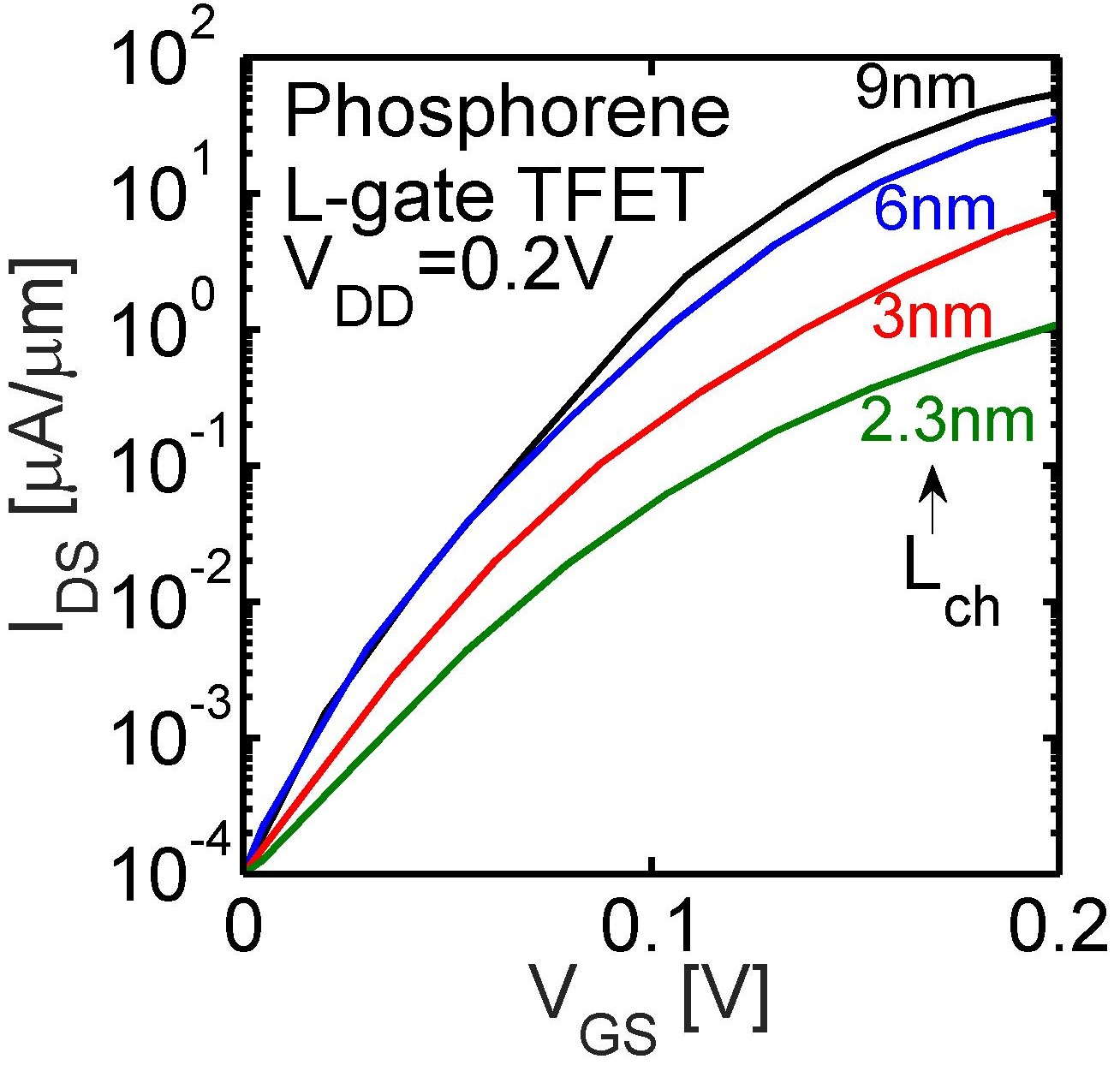} 
               \vspace{-1.5\baselineskip}               
                \caption{}
                \label{fig:laplace}
        \end{subfigure}%
                \begin{subfigure}[b]{0.25\textwidth}
               \includegraphics[width=\textwidth]{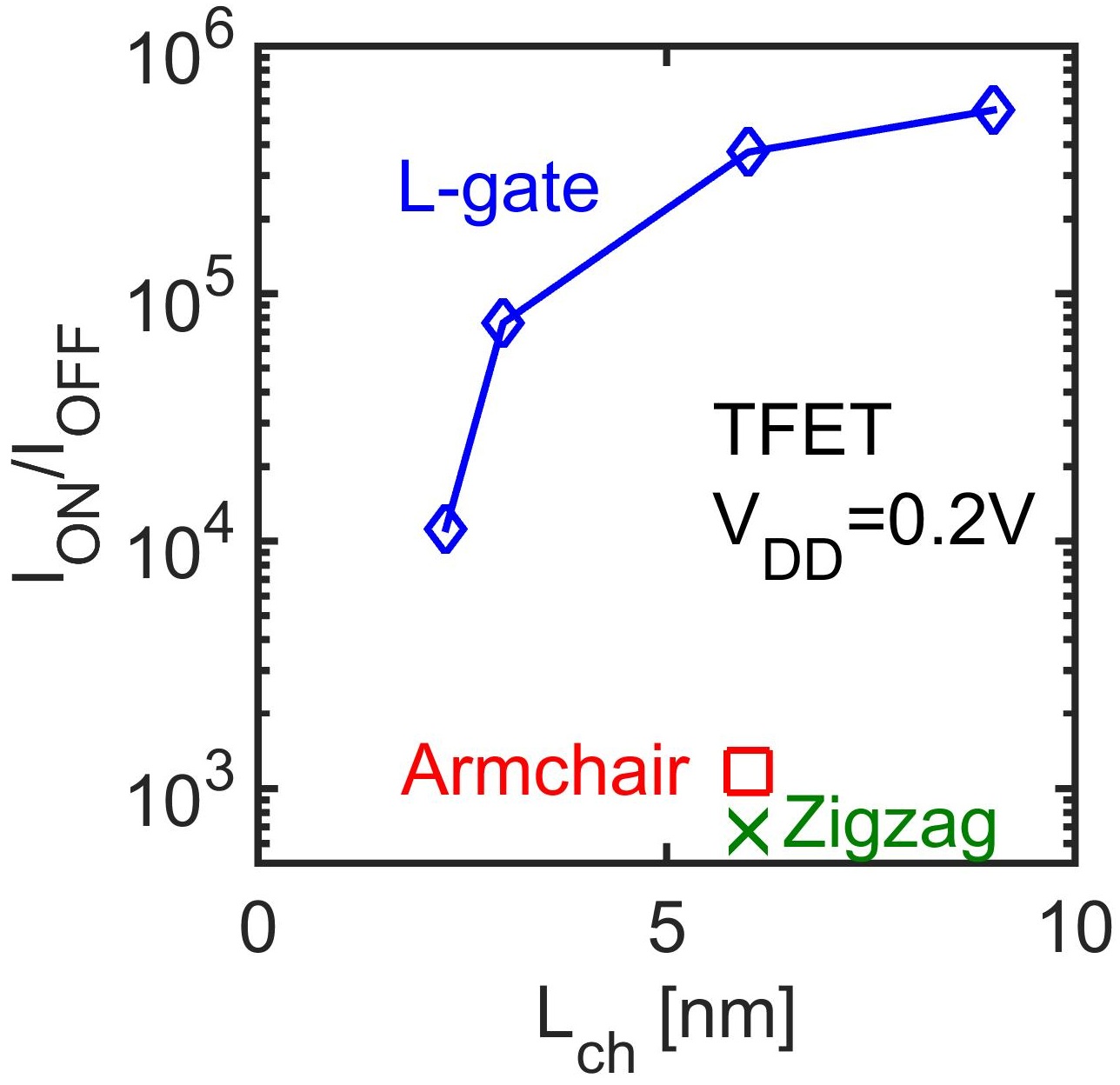} 
               \vspace{-1.5\baselineskip}               
                \caption{}
                \label{fig:laplace}
        \end{subfigure}%
        
        \vspace{-.5\baselineskip}       
        \caption{a) The comparison between $I_D$-$V_G$ of conventional 2L-phosphorene nanoribbons along zigzag and armchair directions with that of the L-gate TFET. b) ON-current of L-gate TFET as a function of $dL$. c) Impact of channel length scaling on $I_D$-$V_G$ of L-gate TFETs. d) I$_{\rm ON}$/I$_{\rm OFF}$ ratio of the L-gate TFET as a function of $L_{ch}$. }\label{fig:Fig3}	
\end{figure}

Fig. \ref{fig:Fig3}a compares $I_D$-$V_G$ of a conventional 2L-phosphorene nanoribbon along zigzag and armchair transport directions with L-shaped gate (L-gate) TFET calculated from full-band atomistic quantum transport simulations using NEMO5 \cite{nemo5_1,nemo5_2}. Not only does the L-gate TFET have I$_{\rm ON}$ close to that of the armchair ribbon (low  $m^*$), but it also has I$_{\rm OFF}$ similar to that of the zigzag ribbon (high $m^*$). Hence, the L-gate design has the advantages of both low and high $m^*$ devices simultaneously: high I$_{\rm ON}$ and low I$_{\rm OFF}$.  

The performance of the L-gate TFET depends on the length $dL$ (see Fig. \ref{fig:Fig3}b) which determines the width for ON-current. In conventional TFETs, $dL$ equals 0. Fig. \ref{fig:Fig3}b shows I$_{\rm ON}$ of L-gate TFET as a function of $dL$ for a fixed I$_{\rm OFF}$ of 10$^{-4} \mu A/\mu m$. Increasing $dL$ enhances I$_{\rm ON}$ significantly, however it reduces the source extension by $dL/2$. Accordingly, there is a limit on $dL$ according to the footage requirements in the design. Nevertheless, a $dL$ of about 2.5nm can improve the performance of TFET approximately by 2 orders of magnitude.

Fig. \ref{fig:Fig3}c shows ultra-scaled L-gate TFETs with channel lengths from 9nm down to 2.3nm with a $V_{DD}$ of 0.2V. In ultra-scaled TFETs, $V_{DD}$ cannot scale below $V_{DD}=0.2V$ since the maximum tunneling energy window is limited by $V_{DD}$. The L-gate TFETs with $L_{ch}$ above 2nm provide I$_{\rm ON}$/I$_{\rm OFF} > 10^4$ and satisfy the minimal ITRS requirement for I$_{\rm ON}$/I$_{\rm OFF}$ ratio. Although the L-gate design has improved the performance of TFETs significantly, the ON-state performance of TFET decreases for devices with $L_{ch}$ and $V_{DD}$ below 2nm and 0.2V, respectively. 

Fig. \ref{fig:Fig3}d shows I$_{\rm ON}$/I$_{\rm OFF}$ ratio of L-gate TFETs as a function of $L_{ch}$. Ultra-scaled channel lengths put a limit on $dL$. Hence, $dL$ shrinks down from 3.5nm to 1nm when the channel length scales down from 9nm to 2.3nm. L-gate TFETs with channel lengths down to 2nm provide I$_{\rm ON}$/I$_{\rm OFF}$ ratio larger than 10$^4$ (required by ITRS as minimum amount of I$_{\rm ON}$/I$_{\rm OFF}$ ratio). This result proves that L-gate TFETs with a channel material of anisotropic $m^*$ enable successful scaling of TFETs down to the ultimate limit; a channel with a few atoms.

\label{sec:opt}

\begin{figure}[!b]
        \centering
        \begin{subfigure}[b]{0.4\textwidth}
               \includegraphics[width=\textwidth]{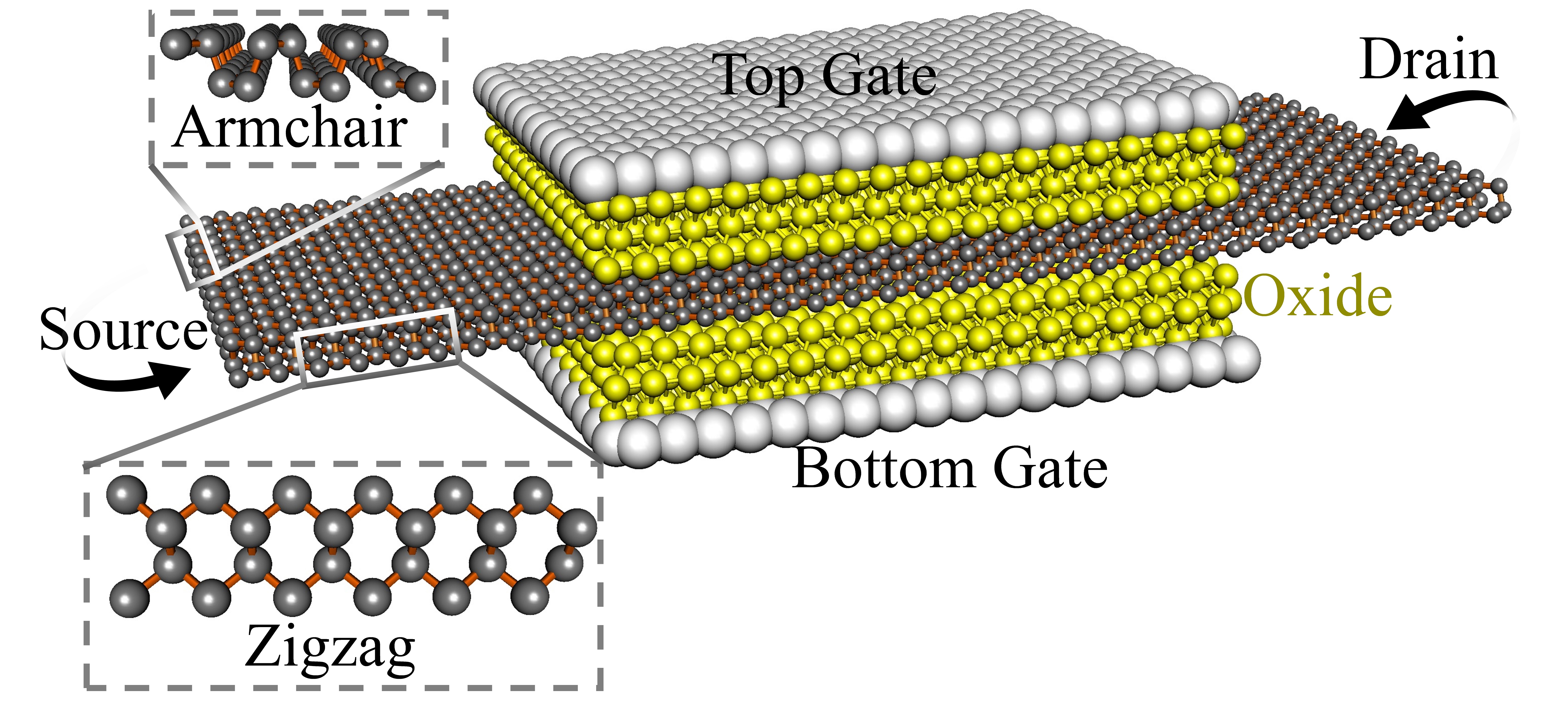}
               \vspace{-.55\baselineskip}               
                \caption{}
                \label{fig:laplace}
        \end{subfigure}%
        \begin{subfigure}[b]{0.27\textwidth}
               \includegraphics[width=\textwidth]{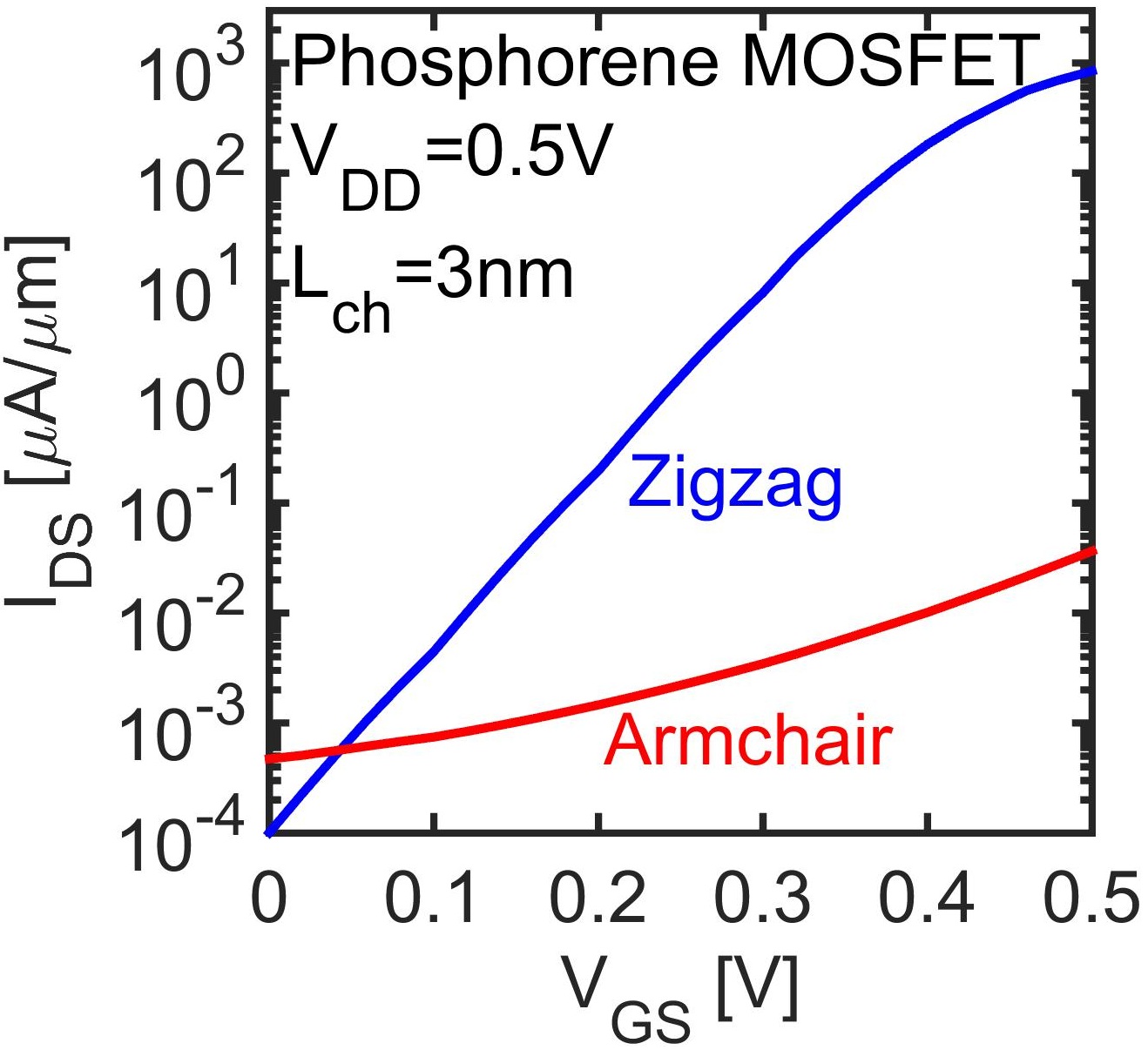} 
               \vspace{-1.5\baselineskip}               
                \caption{}
                \label{fig:laplace}
        \end{subfigure}%
                \quad        
                \begin{subfigure}[b]{0.27\textwidth}
               \includegraphics[width=\textwidth]{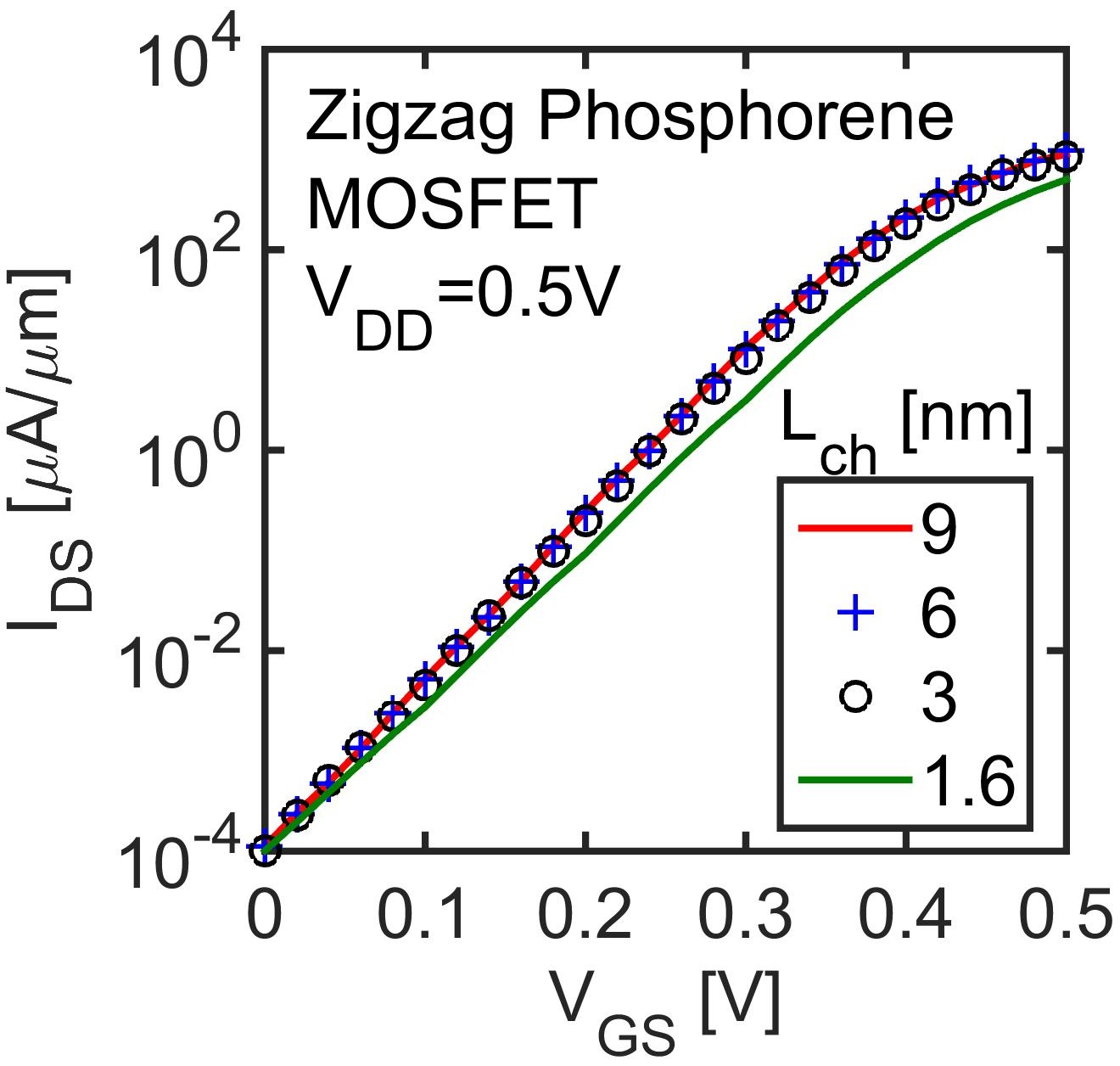} 
               \vspace{-1.5\baselineskip}               
                \caption{}
                \label{fig:laplace}
        \end{subfigure}%
                \begin{subfigure}[b]{0.27\textwidth}
               \includegraphics[width=\textwidth]{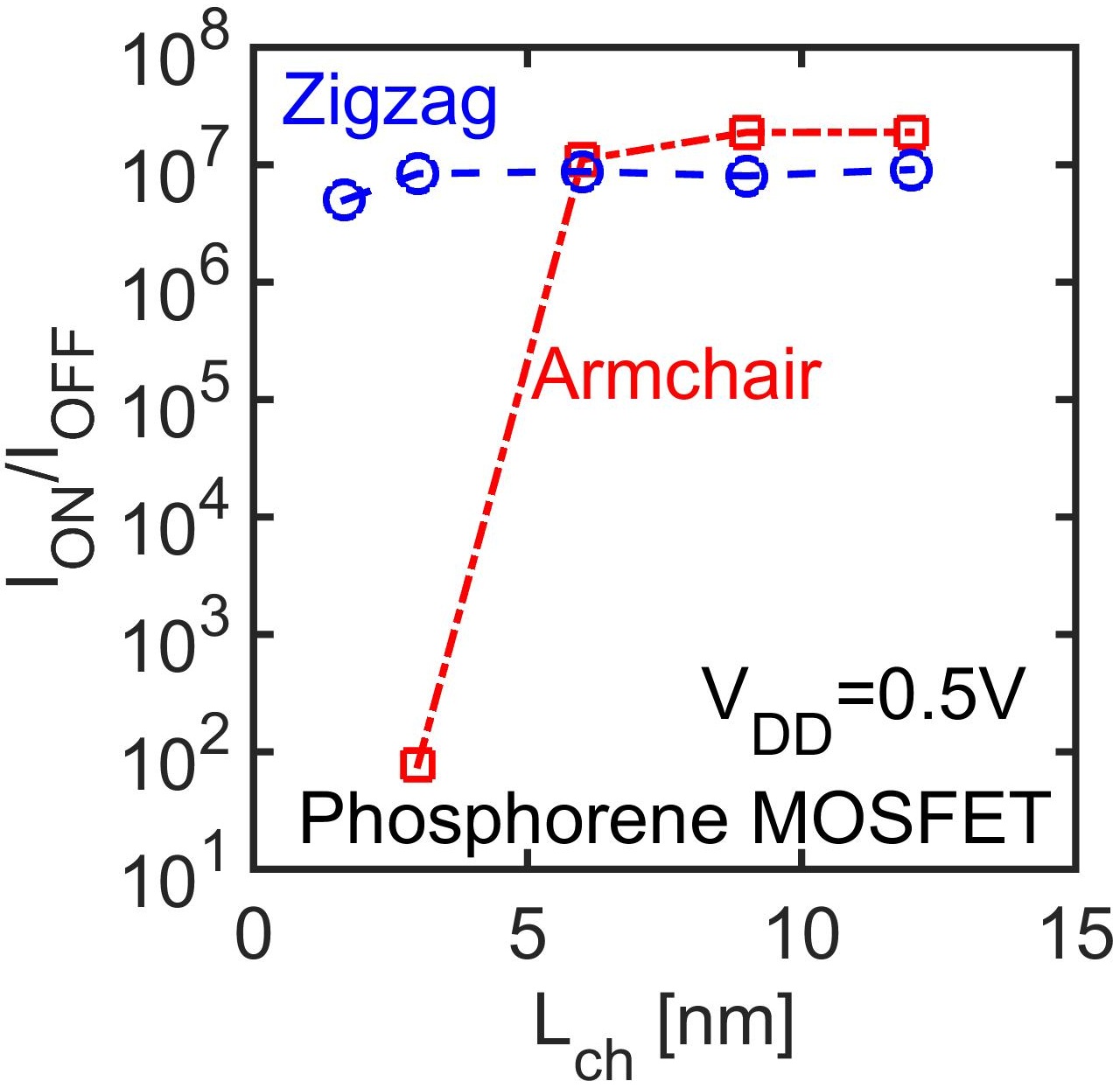} 
               \vspace{-1.5\baselineskip}               
                \caption{}
                \label{fig:laplace}
        \end{subfigure}%
        \vspace{-.5\baselineskip}       
        \caption{a) Device structure of zigzag phosphorene MOSFET. b) The comparison between $I_D$-$V_G$ of phosphorene nanoribbon MOSFETs with transport direction along high $m^*$ (zigzag: blue) and low $m^*$ (armchair: red) axes. c) Impact of $L_{ch}$ scaling on $I_D$-$V_G$ of phosphorene MOSFETs. d) I$_{\rm ON}$/I$_{\rm OFF}$ ratio of MOSFETs as a function of $L_{ch}$ along zigzag and armchair transport directions.}\label{fig:Fig4}	
\end{figure}

As mentioned before, ultra-scaled MOSFETs require large $m^*$ and $E_g$ to block source-to-drain tunneling. Hence, 1L-phosphorene nanoribbon has been chosen here which has the highest $m^*$ and $E_g$ compared to multilayer phosphorene. The schematic of the 1L-phosphorene MOSFET has been shown in Fig. \ref{fig:Fig4}a. The supply voltage is fixed to 0.5V, much higher than $V_{DD}$ of TFETs, since the Boltzmann limit of subthreshold swing in MOSFETs (i.e. 60 mV/decade in room temperature) does not allow the scaling of $V_{DD}$.

The transfer characteristics of a short channel 1L-phosphorene ($L_{ch}$=3nm) with transport direction along low $m^*$ (armchair) and high $m^*$ (zigzag) are compared in Fig. \ref{fig:Fig4}b. As expected, the gate efficiency of a phosphorene MOSFET is much better when the high $m^*$ (zigzag) axis is along the transport direction. This better gate efficiency improves the subthreshold slope of MOSFET significantly. 

Fig. \ref{fig:Fig4}c shows $I_D$-$V_G$ of zigzag scaled phosphorene MOSFETs with channel lengths from 12nm to 1.6nm. Notice that for phosphorene MOSFETs with $L_{ch}>1.6nm$ an $I_{ON}$ larger than $1.1 mA/\mu m$ and an I$_{\rm ON}$/I$_{\rm OFF}$ ratio larger than $10^6$ have been achieved. 1L-phosphorene MOSFETs show a significant advantage over other 2D materials whose performances are diminished below 5nm channel lengths \cite{sub5nm}.

MOSFETs with long channels do not suffer from source-to-drain tunneling. Accordingly, a high transport $m^*$ is not required for blocking this leakage current. Actually, in long channel regime, a low transport $m^*$ can be beneficial and enhance the ON-state performance of the transistor since it leads to a higher carrier injection velocity. Fig. \ref{fig:Fig4}d shows I$_{\rm ON}$/I$_{\rm OFF}$ ratio of phosphorene nanoribbon MOSFETs as a function of $L_{ch}$ along zigzag and armchair transport directions. Although zigzag nanoribbon MOSFETs significantly outperform the armchair ones in short channels due to lower source-to-drain tunneling, armchair nanoribbon MOSFETs show a better performance in longer channels due to higher injection velocity. There is a critical channel length (i.e. 6nm in 1L-phosphorene) in MOSFETs below which having a low $m^*$ becomes critical and above which a high $m^*$ is beneficial.



In summary, the channel materials with anisotropic effective mass can be used to design transistors scalable to 1-2nm channel lengths. In MOSFETs, the high effective mass along transport direction blocks the direct source to drain tunneling and low effective mass reduces the quantum capacitance and switching delay. On the other hand in TFETs, a novel L-shaped gate design is proposed which can provide advantage of high tunneling rate in the ON-state and low tunneling rate in OFF-state by engineering the tunneling paths along low and high effective mass directions. In summary, anisotropic effective mass can be used in an L-gate design to obtain large ON/OFF ratio in an ultra-scaled homojunction TFET.

\section*{Methods}
The atomistic quantum transport simulation results have been obtained from the self consistent solution of 3D-Poisson equation and Non-equilibrium Green's Functions (NEGF) method using the Nanoelectronics modeling tool NEMO5 \cite{nemo5_1, nemo5_2}. The Poisson equation provides the potential for NEGF method and takes the free charge in return. The tight-binding Hamiltonian of phosphorene used in NEGF calculations employs a 10 bands $sp^3d^5s^*$ model. Phosphorene is a material with anisotropic dielectric properties. The details of the Poisson equation with anisotropic dielectric tensor and NEGF equations can be found in our previous works \cite{Hesam1, Tarek1}.

\section*{Acknowledgment}
This work was supported in part by the Center for Low Energy Systems Technology (LEAST), one of six centers of STARnet, a Semiconductor Research Corporation program sponsored by MARCO and DARPA.

\section{Author Contributions}
H. I. came up with the idea of the L-gate TFET. H. I., T. A., and B. N. worked on the atomistic simulations and analyzed the data. Y. T. provided the tight binding model for phosphorene. G. K. and R.R. supervised the work. All authors contributed to writing the manuscript.
\section{Competing financial interests}
The authors declare no competing financial interests.
\end{document}